\documentclass[runningheads]{llncs}
\usepackage{graphicx}
\usepackage{amsmath}
\usepackage{multirow}
\usepackage[thinlines]{easytable}
\usepackage{caption}
\usepackage{subcaption}
\usepackage{amsfonts}
\usepackage[utf8]{inputenc}
\usepackage{amssymb}
\usepackage{calligra}
\usepackage{bbm}
\usepackage[dvipsnames]{xcolor}
\usepackage{enumerate}
\usepackage{paralist}

\usepackage{calrsfs}
\usepackage[mathscr]{euscript}
\DeclareMathAlphabet{\mathcalligra}{T1}{calligra}{m}{n}
\usepackage{calrsfs}
\DeclareMathAlphabet{\pazocal}{OMS}{zplm}{m}{n}
\usepackage[linesnumbered,ruled,vlined]{algorithm2e}
\usepackage{algpseudocode}

\SetKwInput{KwInput}{Input}                
\SetKwInput{KwOutput}{Output}              

\begin{document}
%
\title{Data-driven mapping between functional connectomes using optimal transport}
%
%
 
\author{Javid Dadashkarimi\inst{1} \and
Amin Karbasi\inst{1,2}\and
Dustin Scheinost\inst{3}}
%

%
\institute{Department of Computer Science, Yale University \and
 Department of Electrical Engineering, Computer Science, Statistics \& Data Science, Yale University\\ \and
Department of Radiology and Biomedical Imaging, Yale School of Medicine\\
\email{\{javid.dadashkarimi,amin.karbasi,dustin.scheinost\}@yale.edu}}
\maketitle              
\begin{abstract}
Functional connectomes derived from functional magnetic resonance imaging have long been used to understand the functional organization of the brain. Nevertheless, a connectome is intrinsically linked to the atlas used to create it. In other words, 
a connectome generated from one atlas is different in scale and resolution compared to a connectome generated from  another atlas.
Being able to map connectomes and derived results between different atlases without additional pre-processing
is a crucial step in improving interpretation and generalization between studies that use different atlases.
Here, we use optimal transport, a powerful mathematical technique, to find an optimum mapping between two atlases. This mapping is then used to transform time series from one atlas to another in order to reconstruct a connectome. 
We validate our approach by comparing transformed connectomes against their ``gold-standard'' counterparts (\textit{i.e.}, connectomes generated directly from an atlas) and demonstrate the utility of transformed connectomes by applying  these connectomes to predictive models based on a different atlas.
We show that these transformed connectomes are significantly similar to their ``gold-standard'' counterparts and maintain individual differences in brain-behavior associations, demonstrating both the validity of our approach and its utility in downstream analyses. 
Overall, our approach is a promising avenue to increase the generalization of connectome-based results across different atlases. 

\end{abstract}
\section{Introduction}

Functional connectomics, using functional magnetic resonance imaging (fMRI), are a powerful approach for investigating the functional organization of the brain. 
A prerequisite for creating a functional connectome---\textit{i.e.}, a matrix describing the connectivity between any pair of brain regions---is defining an atlas to parcellate the brain into these regions. 
Given the popularity of this approach, many  atlases, for which there is no gold standard, exist \cite{ARSLAN20185}. 
As these atlases divide the brain into a different number of regions, where each vary by size and topology, connectomes created from different atlases are not directly comparable. 
Thus, results and potential biomarkers generated from one atlas are not readily applicable to connectomes generated from a different atlas.
To extend previous results to a connectome generated from a different atlas, additional preprocessing is needed, a barrier to replication and generalization efforts and limiting wider use of potential connectome-based biomarkers. 

To overcome these limitations, we propose how to find an optimum mapping between two different atlases, allowing data processed from one atlas to be directly transformed into a connectome based on another atlas. 
First, in a training sample with time-series data from two different atlases, we find this mapping by solving the Monge–Kantorovich transportation problem \cite{Peyre:2019}. 
Then, by employing this optimal mapping, time-series data based on the first atlas from novel subjects can be transformed into connectomes based on the second atlas without ever needing to use the second atlas. An overview of our approach is shown in Figure~1.
We validate our approach by comparing transformed connectomes against their ``gold-standard'' counterparts (\textit{i.e.}, connectomes generated directly from an atlas) and demonstrate the utility of transformed connectomes by applying  these connectomes to predictive models based on a different atlas. 
Overall, our results suggest that data from one atlas can be transformed into a connectome comparable to one generated directly from a different atlas.

\begin{figure}[h]
\includegraphics[width=\linewidth]{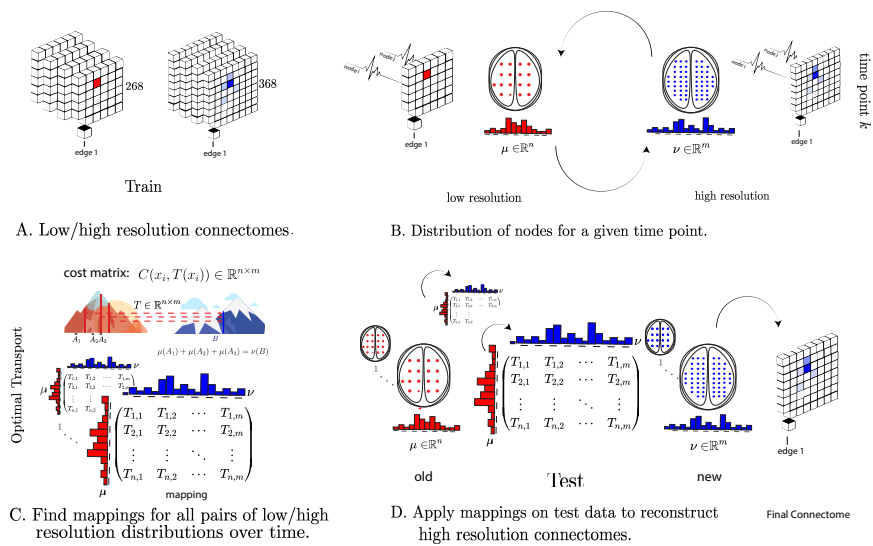}
\caption{Optimal transport pipeline to estimate a mapping between different atlases in order to transform connectomes between the atlases. A) Training data includes  time series data from two atlases, which can be of different number of brain regions. B) Extracting the empirical distribution of brain activity for each node for both atlases at a given time point. C) Learning the optimal transport mapping between source and target distributions for a pre-defined cost matrix. D) Applying the mappings on testing data and then building  transformed connectomes according to the new time series data.  }
\end{figure}

\section{Methods}
\label{Methods}

\subsection{Optimal transport} 
\label{OT}
The optimal transport problem solves how to transport resources from one location $\alpha$ to another $\beta$
while minimizing the cost $C$ to do so \cite{Tolstoi:30,Hitchcock:41,Koopmans:49,Gangbo:96}. 
It has been used for contrast equalization \cite{Delon:04}, image matching \cite{Li:13}, image watermarking \cite{Mathon:14}, text classification \cite{Huang:16}, and music transportation \cite{Flamary:16}.
OT is one of the few methods that
provides a well-defined distance metric when the support of the distributions is different. Other
mappings approaches such as KL divergence do not make this guarantee.


\subsubsection{Monge problem:} 
The original formulation of the optimal transport problem is known as the Monge problem. Lets define some resources $x_1,..,x_n$ in $\alpha$ and some resources $y_1,..,y_m$ in $\beta$. Then, we specify weight vectors $a$ and $b$ over these resources and define matrix $C$ as a measure of pairwise distances between points $x_i \in \alpha$ and comparable points $\pazocal T (x_i)$;
Monge problem aims to solve the following optimizing problem \cite{Monge:81}:
\begin{equation}
    \min_{\pazocal T} \Big\{ \sum_i C(x_i,\pazocal T (x_i)) : \pazocal T_{\sharp} \alpha = \beta \Big\},
\end{equation}
where the push forward operator $\sharp$ indicates that mass from $\alpha$ moves towards $\beta$ assuming that weights absorbed in $b_j = \sum_{\pazocal T(x_i)=y_j} a_i$. 
Assignment problem when the number of elements in the measures are not equal is a special case of this problem, where  each point in $\alpha$ can be assigned to several points in $\beta$.

\subsubsection{Kantorovich relaxation:}
As a generalization of the Monge problem, the Kantorvich relaxation solves the mass transportation problem using a probabilistic approach in which the amount of mass located at $x_i$ potentially dispatches to several points in target \cite{Kantorovich1:42}.  
Admissible solution for Kantorvich relaxation is defined by a coupling matrix $\pazocal{T} \in \mathbb{R}^{n\times m}_+$ indicating the amount of mass being transferred from location $x_i$ to $y_j$ by $\pazocal{T}_{i,j}$:
\begin{equation}
    U(a,b) = \{\pazocal{T} \in \mathbb{R}^{n\times m}_+ : \pazocal{T} \mathbbm{1}_m =a , \pazocal{T}^T\mathbbm{1}_n=b\},
\end{equation}
for vectors of all $1$ shown with $\mathbbm{1}$. An optimum solution is obtained by solving the following problem for a given ``ground metric" matrix $C \in \mathbb{R}^{n\times m}$ \cite{Rubner:2000}:
\begin{equation}
    L_c(a,b) = \min_{\pazocal{T} \in U(a,b)} <C,\pazocal{T}> = \sum_{i,j} C_{i,j} \pazocal{T}_{i,j}.
\end{equation}
which is a linear problem and is not guarantee to have a unique solution \cite{Peyre:2019}, but always there exists an optimal solution (see proof in \cite{Birkhoff:46,Bertsimas:97}).
Kantorovich and Monge problems could also be equivalent in some conditions (see proof in \cite{Brenier:91}).

\subsection{Proposed algorithm for mapping atlases using optimal transport}

\subsubsection{Formulation:}
For paired time-series data from the same individual but from two different atlases (atlas $\kern-4pt\mathcalligra{P_n}\kern1pt$ with $n$ regions and atlas $\kern-4pt\mathcalligra{P_m}\kern1pt$ with $m$ regions), 
lets define $\mu_t \in \mathbb{R}^{n}$ and $\nu_t \in \mathbb{R}^{m}$ to be the distribution of brain activity at single time point $t$ based on atlases $\kern-4pt\mathcalligra{P_n}\kern1pt$ and $\kern-4pt\mathcalligra{P_m}\kern1pt$, respectively.
For a fixed cost matrix $C \in \mathbb{R}^{n\times m}$, we aim to find a mapping $\pazocal{T}\in \mathbb{R}^{n \times m}$ that minimizes transportation cost between $\mu_t$ and $\nu_t$: 
\begin{equation}
    L_c(\mu_t,\nu_t) = \min_{\pazocal{T}}C^T \pazocal{T} \textbf{ s.t, } A\underline{\pazocal{T}}=   \begin{bmatrix}
\mu_t \\
\nu_t 
\end{bmatrix} ,
\end{equation}
in which $\underline{\pazocal{T}}\in \mathbb{R}^{nm}$ is vectorized version of $\pazocal{T}$ such that the $i+n(j-1)$'s element of $\pazocal{T}$ is equal to $\pazocal{T}_{ij}$ and $A$ is defined as:
\begin{equation}
\includegraphics[scale=0.65]{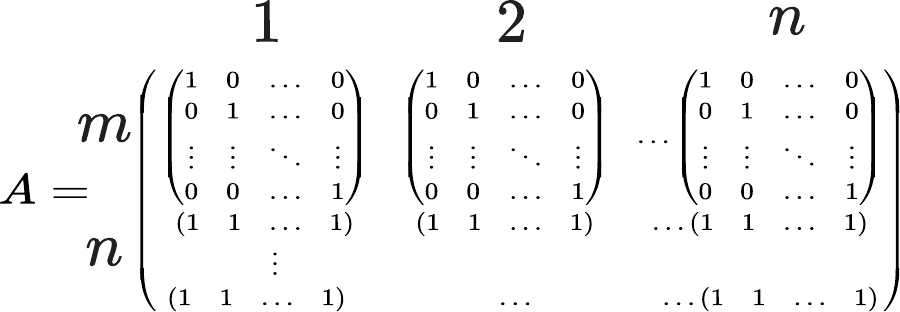}.
\end{equation}
The mapping $\pazocal{T}$ represents the optimal way of transforming the brain activity data from $n$ regions into $m$ regions.

 Yet, solving a large linear program is computationally hard \cite{Dantzig:83}.
Thus, we use the  entropy regularization, which gives an approximation solution with complexity of  $\mathcal{O}(n^2\log (n)\eta^{-3})$ for $\epsilon = \frac{4\log(n)}{\eta}$ \cite{Peyre:2019}, and instead solve the following:
 \begin{equation}
 \label{eq:reg}
    L_c(\mu_t,\nu_t) = \min_{\pazocal{T}}C^T \pazocal{T} - \epsilon H(\pazocal T)\textbf{ s.t, } A\underline{\pazocal{T}}=   \begin{bmatrix}
\mu_t \\
\nu_t 
\end{bmatrix} .
\end{equation}
 Specifically, we use the Sinkhorn algorithm---an iterative solution for Equation~\ref{eq:reg} \cite{Altschuler:2017}---to find the optimum mapping $\pazocal{T}$ as implemented in the Python Optimal Transport (POT) toolbox \cite{flamary2017pot}.

\subsubsection{Defining the cost matrix:}
We investigated two different cost matrices. 
First, we computed the pairwise Euclidean distance between every combination of brain regions between the two atlases by: 
\begin{inparaenum}[\itshape (i\upshape)]
\item computing cluster centroids for all $n$ regions in $\kern-4pt\mathcalligra{P_n}\kern1pt$ and $m$ regions in $\kern-4pt\mathcalligra{P_m}\kern1pt$ and 
\item then, calculating the Euclidean distance between these nodes: $C_{\text{euc}}\left(p,q\right)   = \sqrt {\sum _{i=1}^{3}  \left( q_{i}-p_{i}\right)^2 },$ where $p_i$ and $q_i$ are coordinates for the arbitrary regions: $p$ in $\kern-4pt\mathcalligra{P_n}\kern1pt$ and $q$ in $\kern-4pt\mathcalligra{P_m}\kern1pt$.
\end{inparaenum}
Second, we compute functional distance between regions by: 
\begin{inparaenum}[(i)]
\item calculating the correlation ($\rho$) between the time series for all pairwise combinations of regions between $\kern-4pt\mathcalligra{P_n}\kern1pt$ and  $\kern-4pt\mathcalligra{P_m}\kern1pt$,
\item normalizing by $\rho_{\text{norm}} = (\rho-\min(\rho))/(\max(\rho)-\min (\rho))$, 
\item converting to distance: $C_{\text{func}} = \mathbbm{1}_{n\times m}-\rho_{\text{norm}}$,
\item and average the cost matrix over participants $s$ to get a more robust estimation for $C_{\text{func}}$ (\textit{i.e.}, $C = 1/s \sum_s C_s$). 
\end{inparaenum}

\subsubsection{Estimating mapping:} For training data with $S$ participants and $k$ time points per participant, first, we estimate the optimal mapping $\pazocal T_i$, independently, for each time point and participant using Equation~\ref{eq:reg}. 
The distributions were normalized using
min-max scaling and then dividing by sum.
Next, we 
 average all $\pazocal T_i$ over all participants to produce a single optimal mapping $\pazocal T$ for one time point in the training data (\textit{e.g.}, $\pazocal{T}=\frac{1}{|S|}\sum_{i=1}^{|S|} \pazocal{T}_i$).
 For bigger frames, we use the mapping we learned at the beginning of a frame for the rest time points in the window (i.e., $\pazocal T[t:t+w] \gets  \pazocal T$, for a frame size of $w$ and time point $t$). 

\subsubsection{Estimating connectomes from transformed time series:} Once the $n$ time series from $\kern-4pt\mathcalligra{P_n}\kern1pt$ are transformed to $m$ time series based on  
the target atlas $\kern-4pt\mathcalligra{P_m}\kern1pt$, 
we correlate the time series for every pair of regions $i$ and $j$ to build the final, transformed connectomes.

  
  
        

  

\section{Results}


\subsection{Datasets} 
To evaluate our approach, we used data from the Human Connectome Project (HCP) \cite{Van:2013}, starting with the minimally preprocessed data \cite{Glasser:2013}. 
First, data with a maximum frame-to-frame displacement of 0.15 mm or greater were excluded, resulting in a sample of 876 resting-state scans.
Analyses were restricted only to the LR phase encoding, which consisted of 1200 individual time points. 
Further preprocessing steps were performed using BioImage Suite \cite{Joshi:2011}. 
These included regressing 24 motion parameters, regressing the mean white matter, CSF, and grey matter  time series, removing the linear trend, and low-pass filtering.
Regions were delineated according to the Shen 268 and 368 atlases \cite{Shen:2013}. 
These atlases, defined in an independent dataset, provide a parcellation of the whole gray matter (including subcortex and cerebellum) into 268 or 368 contiguous, functionally coherent regions.
For each scan, the average time series within each region was obtained. To calculate connectomes, the Pearson's correlation between the mean time series of each pair of regions was calculated and converted to be approximately normally distributed using a Fisher transformation.

\begin{table}[t!]
\centering
\begin{tabular}{ccccccccp{0.5cm}cccccc}
&\multicolumn{14}{c}{Train Size (Euclidean Distance)} \\ \cline{2-15}
  &   & \multicolumn{6}{c}{268 $\rightarrow$ 368}                             & \multicolumn{6}{c}{368 $\rightarrow$ 268}                             \\ \cline{2-8} \cline{10-15}
&    & 100    & 200    & 300    & 400    & 500    & 600    &  & 100    & 200    & 300    & 400    & 500    & 600    \\ \cline{2-8} \cline{10-15}
\multirow{11}{*}{\rotatebox[origin=c]{90}{Frame Size}} &100  & 0.489 & 0.495 & 0.491 & 0.498 & 0.496 & 0.494 &  & 0.461 & 0.454 & 0.460 & 0.445 & 0.458 & 0.461 \\
&200  & 0.496 & 0.490 & 0.497 & 0.502 & 0.494 & 0.501 &  & 0.456 & 0.455 & 0.456 & 0.454 & 0.457 & 0.458 \\
&300  & 0.500 & 0.500 & 0.503 & 0.499 & 0.500 & 0.495 &  & 0.447 & 0.454 & 0.451 & 0.454 & 0.453 & 0.458 \\
&400  & 0.490 & 0.499 & 0.492 & 0.496 & 0.499 & 0.494 &  & 0.454 & 0.454 & 0.464 & 0.461 & 0.450 & 0.461 \\
&500  & 0.491 & 0.500 & 0.492 & 0.496 & 0.491 & 0.499 &  & 0.461 & 0.459 & 0.466 & 0.457 & 0.458 & 0.448 \\
&600  & 0.503 & 0.492 & 0.494 & 0.495 & 0.499 & 0.496 &  & 0.454 & 0.455 & 0.456 & 0.457 & 0.454 & 0.452 \\
&700  & 0.491 & 0.508 & 0.492 & 0.500 & 0.493 & 0.498 &  & 0.455 & 0.454 & 0.465 & 0.462 & 0.459 & 0.457 \\
&800  & 0.493 & 0.497 & 0.501 & 0.495 & 0.503 & 0.499 &  & 0.460 & 0.457 & 0.455 & 0.456 & 0.459 & 0.465 \\
&900  & 0.505 & 0.505 & 0.486 & 0.498 & 0.492 & 0.491 &  & 0.451 & 0.456 & 0.460 & 0.461 & 0.462 & 0.458 \\
&1000 & 0.502 & 0.492 & 0.489 & 0.502 & 0.496 & 0.503 &  & 0.452 & 0.457 & 0.469 & 0.450 & 0.461 & 0.452 \\
&1100 & 0.499 & 0.496 & 0.498 & 0.497 & 0.503 & 0.485 &  & 0.453 & 0.460 & 0.455 & 0.460 & 0.453 & 0.464\\ \hline
\end{tabular}
\begin{tabular}{ccccccccp{0.5cm}cccccc}
&\multicolumn{14}{c}{Train Size (Functional Distance)} \\ \cline{2-15}
  &   & \multicolumn{6}{c}{268 $\rightarrow$ 368}                             & \multicolumn{6}{c}{368 $\rightarrow$ 268}                             \\ \cline{2-8} \cline{10-15}
&    & 100    & 200    & 300    & 400    & 500    & 600    &  & 100    & 200    & 300    & 400    & 500    & 600    \\ \cline{2-8} \cline{10-15}
 \multirow{11}{*}{\rotatebox[origin=c]{90}{Frame Size}} & 100  & 0.626 & 0.622 & 0.630 & 0.624 & 0.622 & 0.624 &  & 0.589 & 0.596 & 0.587 & 0.586 & 0.593 & 0.591 \\
&200  & 0.621 & 0.630 & 0.623 & 0.625 & 0.632 & 0.621 &  & 0.592 & 0.597 & 0.590 & 0.601 & 0.584 & 0.591 \\
&300  & 0.624 & 0.629 & 0.629 & 0.627 & 0.629 & 0.630 &  & 0.590 & 0.593 & 0.591 & 0.590 & 0.596 & 0.600 \\
&400  & 0.626 & 0.628 & 0.631 & 0.626 & 0.631 & 0.625 &  & 0.595 & 0.587 & 0.590 & 0.595 & 0.596 & 0.590 \\
&500  & 0.632 & 0.627 & 0.629 & 0.631 & 0.626 & 0.624 &  & 0.594 & 0.598 & 0.596 & 0.601 & 0.591 & 0.593 \\
&600  & 0.633 & 0.630 & 0.631 & 0.631 & 0.629 & 0.631 &  & 0.596 & 0.597 & 0.595 & 0.600 & 0.594 & 0.593 \\
&700  & 0.635 & 0.639 & 0.636 & 0.629 & 0.632 & 0.627 &  & 0.594 & 0.592 & 0.589 & 0.596 & 0.598 & 0.597 \\
&800  & 0.628 & 0.634 & 0.631 & 0.634 & 0.634 & 0.634 &  & 0.598 & 0.596 & 0.590 & 0.592 & 0.593 & 0.601 \\
&900  & 0.632 & 0.633 & 0.634 & 0.635 & 0.635 & 0.638 &  & 0.600 & 0.596 & 0.597 & 0.599 & 0.603 & 0.595 \\
&1000 & 0.632 & 0.635 & 0.635 & 0.638 & 0.643 & 0.636 &  & 0.595 & 0.593 & 0.595 & 0.596 & 0.600 & 0.594 \\
&1100 & 0.638 & 0.636 & 0.638 & 0.634 & 0.639 & 0.639 &  & 0.598 & 0.591 & 0.601 & 0.594 & 0.593 & 0.601 \\ \hline
\end{tabular}
\caption{Intrinsic evaluation of the transformed connectomes based on the optimal mapping $\pazocal T$. The transformed connectomes exhibited high correlation with the ``gold-standard'' connectomes for both (\textit{top}) the Euclidean distance and (\textit{bottom}) the functional distance cost matrices. Similarity between connectomes was not affected by sample size and number of time points used to estimate $\pazocal T_i$.}
\label{tab:results}
\end{table}

\subsection{Intrinsic evaluation}
\label{Intrinsic Evaluation}
\subsubsection{Correlation with ``gold-standard'' connectomes:}
To validate our approach, 
we, first, partitioned our sample into 80\% training data to estimate the optimal mapping $\pazocal T$ between atlases and 20\% testing data for evaluating the quality of the transformed connectomes. 
In the training data, we estimated $\pazocal T$  using all 1200 time points and 700 participants for each of the cost matrices ($C_{\text{euc}}$ and $C_{\text{func}}$). Next, in the testing data, we applied $\pazocal T$ to construct $368 \times 368$ connectomes from the 268 atlas data (labeled: $268 \rightarrow 368$) as well as $268 \times 268$ connectomes from the 368 atlas data (labeled: $368 \rightarrow 268$). Finally, the transformed connectomes were compared to the ``gold-standard'' connectomes (\textit{i.e.}, connectomes generated directly from an atlas) using correlation. Using a 12 core processor Intel Xeon Gold 6128 CPU with a 3.40GHz clock speed, estimating $\pazocal T$ took  2,975s.

For both cost matrices, significant correlations between the transformed connectomes and the ``gold-standard'' connectomes were observed (for $C_{\text{euc}}$, $268 \rightarrow 368$: $r=0.508, p<0.01$;  $368 \rightarrow 268$: $r=0.469,p<0.01$; for $C_{\text{func}}$, $268 \rightarrow 368$: $r=0.643,p<0.01$;  $368 \rightarrow 268$: $r=0.603,p<0.01$). Notably, transformed connectomes using the $C_{\text{func}}$ cost matrix were significantly ($p<0.01$) more similar to the ``gold-standard'' connectomes compared to transformed connectomes using the $C_{\text{euc}}$ cost matrix. Finally, the $268 \rightarrow 368$ connectomes were more similar to the ``gold-standard'' connectomes compared to  the $368 \rightarrow 268$ connectomes. 

\subsubsection{Evaluation of free parameters:}
Next, we investigated the sensitivity of our approach to the number of time points and number of participants used to find the mapping between atlases. Using the $80/20$ split for training and testing, we varied the number of time points used from $100$ to $1100$ in $100$ increments and varied the number of participants from $100$ to $600$ in $100$ increments. 
No clear pattern of performance change was observed across either parameter,
suggesting that our approach is stable to both the number of frames and participants (Table~\ref{tab:results}).
However, using only $100$ participants and $100$ time points \textcolor{black}{in a frame} significant ($p<0.05$) reduced the 
processing time from 2,975 s to 467s.

\subsection{Extrinsic evaluation}
\label{Extrinsic Evaluation}
In addition to validating our approach, we demonstrated that the transformed connectomes can be used to elucidate brain-behavior associations. 
To this aim, 
\begin{inparaenum}[\bfseries 1\upshape)]
\item We partitioned our data into three folds $g_1$, $g_2$, and $g_3$ with a respective ratio of $\{0.25,0.5,0.25\}$. 
\item  Using only participants in $g_1$, we estimated the optimal mapping $\pazocal T$ for both cost matrices. 
\item We applied $\pazocal T$ to the participants in $g_3$ to produce the transformed connectomes ($268 \rightarrow 368$ and $368 \rightarrow 268$).
\item We predicted IQ using ridge regression \cite{Gao:2019} and classified sex using support vector machine (SVM) with a linear kernel \cite{Cortes:1995} using the connectomes in $g_2$ for both the $268 \time 268$ and $368 \time 368$, independently. All models were trained with $10$-fold cross-validation.
\item We used the predictive models from Step \textbf{4)} to
predict phenotypic information using the transformed matrices from Step \textbf{3)} (\textit{e.g.}, using the $268 \rightarrow 368$ connectomes as inputs to the models trained with the $368 \time 368$ connectomes).
\end{inparaenum}
We tested the significance of predictions based on the transformed connectomes against a null distribution of prediction based on permuted values using corrected resampled t-tests \cite{10.1007/978-3-540-24775-3_3}. 

\begin{figure*}[t]
    \centering
    \begin{subfigure}[t]{0.5\textwidth}
        \centering
        \includegraphics[width=\textwidth]{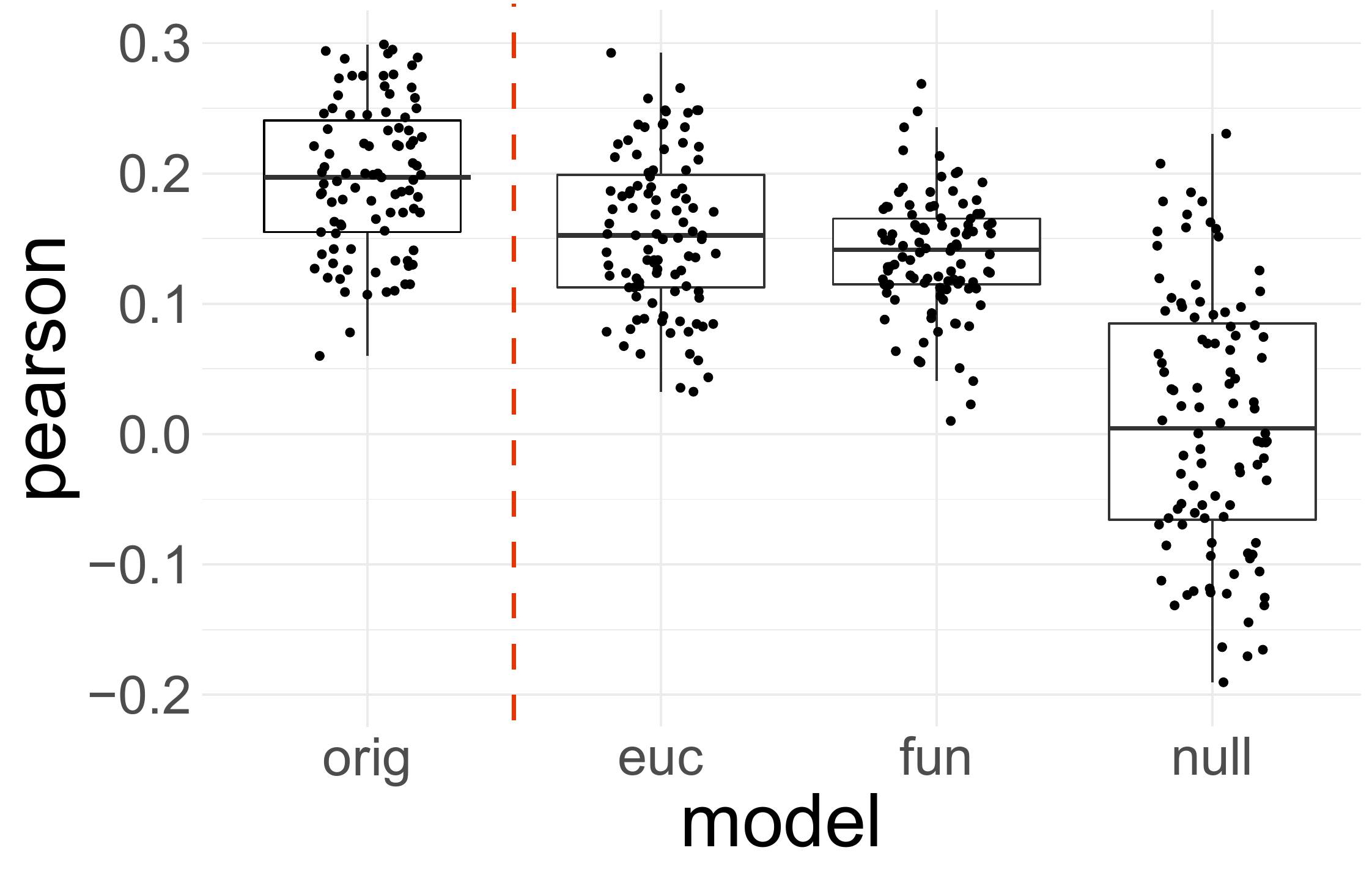}
        \caption{268 $\rightarrow$ 368}
        \label{fig:iq-left}
    \end{subfigure}%
    \begin{subfigure}[t]{0.5\textwidth}
        \centering
        \includegraphics[width=\textwidth]{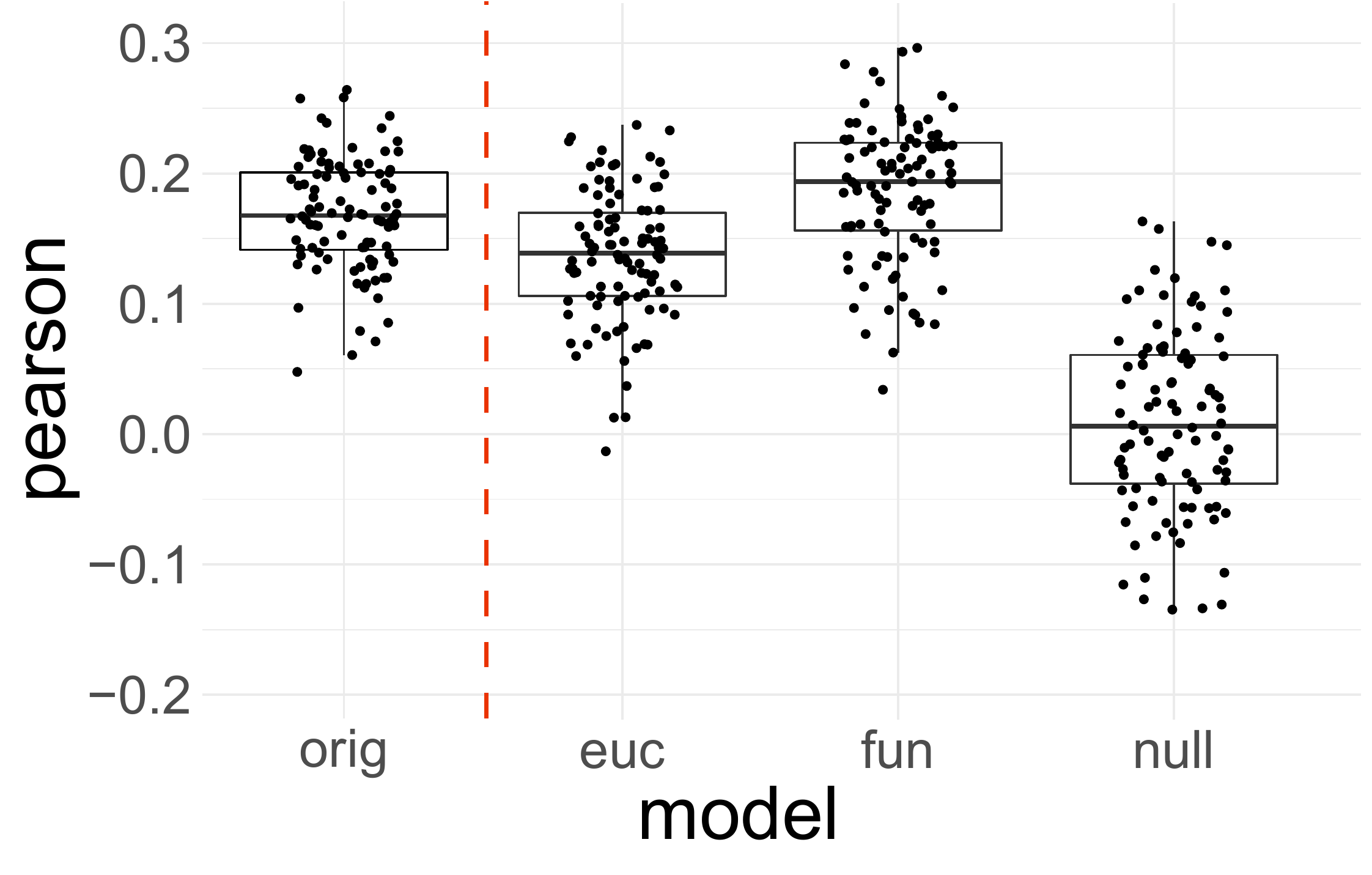}
        \caption{368 $\rightarrow$ 268}
        \label{fig:iq-right}
    \end{subfigure}%
    \caption{Box plots for IQ prediction from transformed connectomes. Participants were randomly split into three groups ($g_1$, $g_2$, and $g_3$) with a respective ratio of $\{0.25,0.5,0.25\}$. A mapping is trained on $g_1$, the model is trained on connectomes from $g_2$, and tested on transformed connectomes from $g_3$. \textit{orig} shows prediction performance in ``gold-standard'' connectomes, \textit{euc} and \textit{fun} show prediction performance for transformed connectomes found using either $C_{\text{euc}}$ or $C_{\text{func}}$, respectively. 
    A null model (labeled \textit{null}) is obtained by permuting labels.}
    \label{fig:iq}
\end{figure*}

\begin{table}[t]
    \centering
\begin{tabular}{c c |c|c|c | c | c c|c|c|} \cline{3-10}
&\multicolumn{5}{c}{sex} & & \multicolumn{3}{c}{iq} \\ 
\cline{3-5} \cline{7-10}
 & & null      & euc    & func   && & null      & euc    & func   \\ \cline{3-5} \cline{7-10}
\multirow{1}{*}{268 $\rightarrow$ 368} && 0.5033  & 0.6961$^*$ & 0.7253$^*$ & && 0.0083  & 0.1553$^*$ & 0.1376$^*$ \\ \cline{3-5} \cline{7-10}
\multirow{1}{*}{368 $\rightarrow$ 268}  && 0.5077  & 0.7312$^*$ & 0.7243$^*$ &  & & 0.0036  & 0.1313$^*$ & 0.1835$^*$ \\ \cline{3-5} \cline{7-10}
\end{tabular}
\caption{Extrinsic evaluation of the transformed connectomes based on the optimal mapping $\pazocal T$. Indicator $^*$ shows the significance of the results with respect to the null model for $p<0.05$ using corrected resampled t-tests \cite{10.1007/978-3-540-24775-3_3}. }
\label{tab:results2}
\end{table}

Results showed that using transformed connectomes from both cost matrices and both directions (\textit{e.g.}, $368 \rightarrow 268$)
lead to significantly ($p<0.05$) better prediction of IQ compared to the null model (see Table~\ref{tab:results2}, Figure~\ref{fig:iq-left}).
Similarly, results showed that sex classification achieves up to 72\%  accuracy and is significantly higher compared to the null distribution for all transformed connectomes (see Table~\ref{tab:results2}). There were no clear differences in prediction performance between the transformed connectomes from either cost matrix or direction. 

Finally, we compared the prediction performance of the ``gold-standard'' connectomes and the transformed connectomes.
In this comparison, the prediction performance of the ``gold-standard'' connectomes was treated 
as an upper limit of how well the transformed connectomes could perform, as it is unreasonable to expect the transformed connectomes to outperform the `gold-standard'' connectomes.
For both IQ and sex prediction, prediction performance of the transformed connectomes overlapped that of the ``gold-standard'' connectomes, indicating that little information about brain-behavior association is lost when transforming data using the estimated optimal mapping $\pazocal T$. 



\section{Discussion and conclusions}
Atlas selection is a prerequisite for creation of a functional connectome. Yet, any choice of atlas ultimately constrains interpretation and future replication and generalization efforts to that particular atlas. 
Since there is no single gold-standard atlas, results generated from two distinct atlases must undergo additional processing before comparison.
In this work, we propose optimal transport to find optimum mappings between different atlases, which enable data, previously processed with one atlas, to be mapped to a connectome generated from a different atlas, without the need for further prepossessing. We show that these transformed connectomes are significantly similar to their ``gold-standard'' counterparts and maintain individual differences in brain-behavior associations, demonstrating both the validity of our approach and its utility in downstream analyses. 
Ourapproach is in the spatial domain (i.e., mapping node-to-node), rather than mapping
timecourse-to-timecourse or even connectome-to-connectome. Once we have a node-to-node
mapping, timecourses (and resulting connectomes) naturally come for free. While our end goal
is to generate the transformed connectomes, we chose the node-to-node approach as it is more
general. 
Importantly, our optimal mappings are robust to training parameters, suggesting that a single mapping between any atlas pair could be generated once and used as an off-the-shelf solution by the community. Future work will include further validation of our approach in a wider range of atlases and the generation of a publicly available repository of mappings for community use. Overall, our approach is a promising avenue to increase the generalization of connectome-based results across different atlases. 

\subsubsection{Acknowledgements:}
Data were provided in part by the Human Connectome Project, WU-Minn Consortium (Principal Investigators: David Van Essen and Kamil Ugurbil; U54 MH091657) and funded by the 16 NIH Institutes and Centers that support the NIH Blueprint for Neuroscience Research; and by the McDonnell Center for Systems Neuroscience at Washington University. We thank Brendan Adkinson for the helpful comments on this work.

%
%
%
%
\bibliographystyle{splncs04}
\bibliography{sample}

\begin{thebibliography}{10}
\providecommand{\url}[1]{\texttt{#1}}
\providecommand{\urlprefix}{URL }
\providecommand{\doi}[1]{https://doi.org/#1}

\bibitem{Altschuler:2017}
Altschuler, J., Weed, J., Rigollet, P.: Near-linear time approximation
  algorithms for optimal transport via sinkhorn iteration. arXiv preprint
  arXiv:1705.09634  (2017)

\bibitem{ARSLAN20185}
Arslan, S., Ktena, S.I., Makropoulos, A., Robinson, E.C., Rueckert, D.,
  Parisot, S.: Human brain mapping: A systematic comparison of parcellation
  methods for the human cerebral cortex. NeuroImage  \textbf{170},  5--30
  (2018). \doi{https://doi.org/10.1016/j.neuroimage.2017.04.014},
  \url{https://www.sciencedirect.com/science/article/pii/S1053811917303026},
  segmenting the Brain

\bibitem{Bertsimas:97}
Bertsimas, D., Tsitsiklis, J.: Introduction to linear optimization, athena
  scientific, 1997. URL: http://athenasc. com/linoptbook. html

\bibitem{Birkhoff:46}
Birkhoff, G.: Tres observaciones sobre el algebra lineal. Univ. Nac. Tucuman,
  Ser. A  \textbf{5},  147--154 (1946)

\bibitem{10.1007/978-3-540-24775-3_3}
Bouckaert, R.R., Frank, E.: Evaluating the replicability of significance tests
  for comparing learning algorithms. In: Dai, H., Srikant, R., Zhang, C. (eds.)
  Advances in Knowledge Discovery and Data Mining. pp. 3--12. Springer Berlin
  Heidelberg, Berlin, Heidelberg (2004)

\bibitem{Brenier:91}
Brenier, Y.: Polar factorization and monotone rearrangement of vector-valued
  functions. Communications on pure and applied mathematics  \textbf{44}(4),
  375--417 (1991)

\bibitem{Cortes:1995}
Cortes, C., Vapnik, V.: Support-vector networks. Machine learning
  \textbf{20}(3),  273--297 (1995)

\bibitem{Dantzig:83}
Dantzig, G.B.: Reminiscences about the origins of linear programming. In:
  Mathematical Programming The State of the Art, pp. 78--86. Springer (1983)

\bibitem{Delon:04}
Delon, J.: Midway image equalization. Journal of Mathematical Imaging and
  Vision  \textbf{21}(2),  119--134 (2004)

\bibitem{flamary2017pot}
Flamary, R., Courty, N.: Pot python optimal transport library (2017),
  \url{https://pythonot.github.io/}

\bibitem{Flamary:16}
Flamary, R., F{\'e}votte, C., Courty, N., Emiya, V.: Optimal spectral
  transportation with application to music transcription. arXiv preprint
  arXiv:1609.09799  (2016)

\bibitem{Gangbo:96}
Gangbo, W., McCann, R.J.: The geometry of optimal transportation. Acta
  Mathematica  \textbf{177}(2),  113--161 (1996)

\bibitem{Gao:2019}
Gao, S., Greene, A., Constable, T., Scheinost, D.: Combining multiple
  connectomes improves predictive modeling of phenotypic measures. Neuroimage
  \textbf{In Press} (2019)

\bibitem{Glasser:2013}
Glasser, M.F., Sotiropoulos, S.N., Wilson, J.A., Coalson, T.S., Fischl, B.,
  Andersson, J.L., Xu, J., Jbabdi, S., Webster, M., Polimeni, J.R., et~al.: The
  minimal preprocessing pipelines for the human connectome project. Neuroimage
  \textbf{80},  105--124 (2013)

\bibitem{Hitchcock:41}
Hitchcock, F.L.: The distribution of a product from several sources to numerous
  localities. Journal of mathematics and physics  \textbf{20}(1-4),  224--230
  (1941)

\bibitem{Huang:16}
Huang, G., Quo, C., Kusner, M.J., Sun, Y., Weinberger, K.Q., Sha, F.:
  Supervised word mover's distance. In: Proceedings of the 30th International
  Conference on Neural Information Processing Systems. pp. 4869--4877 (2016)

\bibitem{Joshi:2011}
Joshi, A., Scheinost, D., Okuda, H., Belhachemi, D., Murphy, I., Staib, L.H.,
  Papademetris, X.: Unified framework for development, deployment and robust
  testing of neuroimaging algorithms. Neuroinformatics  \textbf{9}(1),  69--84
  (2011)

\bibitem{Kantorovich1:42}
Kantorovich, L.: On the transfer of masses (in russian). In: Doklady Akademii
  Nauk. vol.~37, pp. 227--229 (1942)

\bibitem{Koopmans:49}
Koopmans, T.C.: Optimum utilization of the transportation system. Econometrica:
  Journal of the Econometric Society pp. 136--146 (1949)

\bibitem{Li:13}
Li, P., Wang, Q., Zhang, L.: A novel earth mover's distance methodology for
  image matching with gaussian mixture models. In: Proceedings of the IEEE
  International Conference on Computer Vision. pp. 1689--1696 (2013)

\bibitem{Mathon:14}
Mathon, B., Cayre, F., Bas, P., Macq, B.: Optimal transport for secure
  spread-spectrum watermarking of still images. IEEE Transactions on Image
  Processing  \textbf{23}(4),  1694--1705 (2014)

\bibitem{Monge:81}
Monge, G.: M{\'e}moire sur la th{\'e}orie des d{\'e}blais et des remblais.
  Histoire de l'Acad{\'e}mie Royale des Sciences de Paris  (1781)

\bibitem{Peyre:2019}
Peyr{\'e}, G., Cuturi, M., et~al.: Computational optimal transport: With
  applications to data science. Foundations and Trends{\textregistered} in
  Machine Learning  \textbf{11}(5-6),  355--607 (2019)

\bibitem{Rubner:2000}
Rubner, Y., Tomasi, C., Guibas, L.J.: The earth mover's distance as a metric
  for image retrieval. International journal of computer vision
  \textbf{40}(2),  99--121 (2000)

\bibitem{Shen:2013}
Shen, X., Tokoglu, F., Papademetris, X., Constable, R.T.: Groupwise whole-brain
  parcellation from resting-state fmri data for network node identification.
  Neuroimage  \textbf{82},  403--415 (2013)

\bibitem{Tolstoi:30}
Tolstoi, A.: Methods of finding the minimal total kilometrage in cargo
  transportation planning in space. TransPress of the National Commissariat of
  Transportation  \textbf{1},  23--55 (1930)

\bibitem{Van:2013}
Van~Essen, D.C., Smith, S.M., Barch, D.M., Behrens, T.E., Yacoub, E., Ugurbil,
  K., Consortium, W.M.H., et~al.: The wu-minn human connectome project: an
  overview. Neuroimage  \textbf{80},  62--79 (2013)

\end{thebibliography}

\end{document}